\documentclass[12pt]{article}
\usepackage{graphics,color}
\begin{document}
\thispagestyle{empty}
\noindent\
\\
\\
\\
\begin{center}
\large \bf Composite Weak Bosons
\end{center}
\hfill
 \vspace*{1cm}
\noindent
\begin{center}
{\bf Harald Fritzsch}\\
Department f\"ur Physik, Universit\"at M\"unchen,\\
Theresienstra{\ss}e 37, 80333 M\"unchen
\vspace*{0.5cm}
\end{center}

\begin{abstract}
The weak bosons, leptons and quarks are considered as
composite particles. The interaction of the constituents
is a confining gauge interaction. The standard
electroweak model is a low energy
approximation. The mixing of the neutral weak boson with the photon
is a dynamical mechanism, similar to the mixing between the photon
and the $\rho$-meson in QCD. This mixing provides information
about the energy scale of the confining gauge force. It must be
less than 1 TeV.
At and above this energy
many narrow resonances should exist, which decay
into weak bosons and into lepton and quark pairs. Above 1 TeV excited
leptons should exist, which decay into leptons under emission of
a weak boson or a photon. These new states can be observed
with the detectors at the Large Hadron Collider in CERN.
\end{abstract}

\newpage

In the Standard Model of the electroweak interactions,
based on the gauge group
$SU(2) \times U(1)$ (ref. (1)), the universality of the
weak interactions follows from the electroweak gauge symmetry. The
masses of the weak bosons and of the leptons and quarks are
introduced by the spontaneous breaking of the gauge symmetry (ref.(2)).\\

The masses of the bound states in confining gauge
theories are generated by another mechanism, which follows from the confinement
property of a non-Abelean gauge force. In QCD the masses
of the hadrons in the absence of the
quark masses are given by the field energy of the confined gluons
and quarks. The masses are proportional to the QCD
mass scale $\Lambda_c$, which has been measured in deep inelastic
scattering to about 300 MeV.\\

If the weak bosons, the leptons and quarks are composite particles, their masses should
also be due to the confinement mechanism. The
constituents are confined by a gauge force analogous to QCD.
The fundamental constituents are elementary fermions
and bosons, denoted here as "haplons" (derived
from "haplos", the Greek translation of "simple").
The confining gauge group is unknown - it might be
$SU(n)$, and the number n might be 3, as in QCD. Such a
theory, denoted as QHD (H stands for "haplon") would replace the standard electroweak gauge theory.
Scalar fields are not needed for the generation of the masses of the leptons, quarks and
weak bosons. In 1981 we studied a specific composite model of this type (ref.(3), see also
refs.(4,5,6,7,8)). \\

If the weak bosons are composite, the weak interaction is not
a fundamental gauge interaction as in the Standard Model, but an
effective interaction, analogous to the nuclear force. The
weak bosons are not fundamental gauge particles, but composite
objects with a finite radius. Their masses are
given by the field energy of the gauge bosons and of the
basic constituents, confined inside the weak bosons.\\

We assume that the weak bosons are bound states of two elementary
fermions, as the vector mesons in QCD. The weak currents are bilinear in the fermion fields.
The weak bosons are QHD singlets.\\

The fundamental fermions are lefthanded.
The lowest QHD singlets are vector particles, consisting of a
haplon and its antiparticle.
The
intrinsic mass scale of the QHD gauge interaction is
denoted by $\Lambda_h$. The length scale,
associated to this mass scale, defines the radius of the weak bosons.\\

QHD is a chiral gauge theory. This implies a maximal violation of parity
in the weak interactions.
The doublet of the weak isospin group $SU(2)$ is given by two lefthanded haplon fields h:
\begin{equation}
h = \left( \begin{array}{l}
\alpha\\
\beta\\
\end{array} \right) \; .
\end{equation}
Their electric charges in units of e are:
\begin{eqnarray}
Q(\alpha) & = & +1/2 \; , \nonumber \\
Q(\beta) & = & -1/2 \; .
\end{eqnarray}
The three
W-bosons have the following internal structure:
\begin{eqnarray}
W^+ & = & \overline{\beta} \alpha \; , \nonumber \\
W^- & = & \overline{\alpha} \beta \; , \nonumber \\
W^3 & = & \frac{1}{\sqrt{2}} \left( \overline{\alpha} \alpha -
\overline{\beta} \beta \right) \; .
\end{eqnarray}
\\
The leptons and quarks are bound states of a fermion and a scalar
haplon, transforming under the QHD group as the fundamental representation.
There are four scalars - one for leptons and three for the
three colors of the quarks. The electric charges of the scalars are
-1/2 e for the lepton scalar and +1/6 e for a quark scalar (details: see
ref. (3)). If the mass scale $\Lambda_h$ is about 1 TeV, the
leptons and quarks and the weak bosons would have a finite extension
of about $10^{-17}$ cm, just below the experimental limit.\\

As in QCD the spectral functions of
the weak currents would be dominated at energies below the
mass scale $\Lambda_h$ by the lowest vector bosons,
the weak bosons. At the energy of the order of
$\Lambda_h$ exist many new narrow resonances, including the excited
states of the W- bosons.\\

The weak bosons couple universally to the lepton and quark doublets.
The weak currents obey the local isospin
current algebra. We deduce the universality of the weak couplings from the
dominance of the weak currents by the lowest W-bosons (ref. (5)). Thus
the universality does not imply an underlying gauge principle. In
QCD the universality of the couplings of the
vector mesons to the hadrons follows analogously from the current
algebra and the dominance of the matrix elements of the vector
current by the lowest vector meson pole.\\

The dynamics of the chiral gauge theory QHD must be different from QCD.
 In QCD the pions are lower in mass than $\Lambda_c$, but the vector mesons
have masses above $\Lambda_c$. In QHD the weak bosons are significantly lower
in mass than the QHD mass scale. This must be due to the chiral structure of
QHD. Details of the bound state structure of chiral confining gauge
theories are not yet known.\\

In QCD the pions are massless in the limit, where the u- and d-quark masses
vanish. The charged pions will have a small mass, due to the electromagnetic self energy.
In QHD the leptons and quarks are assumed to be massless, if the electromagnetic and weak interactions are turned off.
Their masses are due to electroweak corrections and might be calculable in the future. \\

We proceed to estimate the mass scale of the QHD interaction.
We consider the three $\rho$-mesons in QCD. If the electromagnetic
interaction is turned off, the three
$\rho$-mesons are degenerate in mass. This degeneracy is lifted, when
the electromagnetism is turned on. The charged $\rho$-mesons aquire a
mass due to the Coulomb repulsion of the quarks. The photon and the neutral
meson mix, and the neutral meson receives a mass
contribution. The mass shift between the charged and neutral
$\rho$-mesons can be calculated. It depends on the
decay constant of the $\rho$-meson and on the electromagnetic coupling
constant. The mixing parameter m is given by:
\begin{equation}
m \; =\; e \frac{F^{}_\rho}{M^{}_\rho} \; .
\end{equation}
Experimentally the mixing parameter m is about 0.09, and it leads to a mass
shift between the neutral and charged $\rho$-meson of about 1.2 MeV.\\

Analogously in QHD in the absence of electromagnetism the three weak
bosons are degenerate in mass, due to the flavor symmetry of the two
haplons (1). When the electromagnetic interaction is turned on, the
photon and the neutral weak boson mix. This mixing causes a mass shift
for the Z-boson, which in the experiments has been measured to 10.79 +/- 0.03 GeV.\\

In the Standard Model this
mixing is generated by the scalar fields and is described by the
weak mixing angle. In QHD the mixing is a dynamical phenomenon, analogous to the
mixing of the photon and the vector meson in QCD. The associated mixing parameter is
given by the decay constant of the
W-boson, defined in analogy to the decay constant of the vector meson in QCD:
\begin{equation}
\langle 0 \left| \frac{1}{2} \left( \overline{\alpha} \gamma^{}_\mu
\alpha - \overline{\beta} \gamma^{}_\mu \beta \right) \right| Z
\rangle \; =\; \varepsilon^{}_\mu M^{}_W F^{}_W \; .
\end{equation}
The mixing parameter m is given again by the decay constant:
\begin{equation}
m \; =\; e \frac{F^{}_W}{M^{}_W} \; .
\end{equation}
In the Standard Model m is given by the weak angle:
\begin{equation}
\sin\theta^{}_w \; =\; m \;.
\end{equation}
The mass difference between the Z-boson  and the W-boson is
determined by the mixing parameter m and the W-mass:
\begin{equation}
M^2_Z - M^2_W \; =\; M^2_W \left(\frac{m^2}{1 - m^2}\right) \; .
\end{equation}
We shall use the following experimental values:
\begin{eqnarray}
M^{}_W & = & 80.4 ~{\rm GeV} \; , \nonumber \\
M^{}_Z & = & 91.9 ~{\rm GeV} \; , \nonumber \\
F^{}_W & = & 124 ~{\rm GeV} \; , \nonumber \\
\sin^2 \theta^{}_W & = & 0.2315 \; , \nonumber \\
\alpha & = & \frac{e^2}{4\pi} \; \approx \; \frac{1}{128.9} \; ,
\nonumber \\
e & = & 0.3122 \; , \nonumber \\
m & = & 0.482 \; .
\end{eqnarray}
In QCD there is a direct connection between the decay constant of
the vector meson and the QCD mass parameter $\Lambda_c$. The decay
constant of the vector mesons is measured to about 220 MeV. The QCD
mass parameter $\Lambda_c$ has been measured to about 350 MeV. The ratio of the
QCD mass parameter and the decay constant is about 1.6.\\

There should be a similar connection between the decay
constant of the weak boson and the QHD mass parameter $\Lambda_h$.
If the QHD gauge group would be SU(3), we would expect
that the ratio of $\Lambda_h$ and $\Lambda_c$ is given by the measured
ratio of the decay constants, i.e. 560, and $\Lambda_h$ would be about 0.20 TeV.
For other groups $\Lambda_h$
could be larger, but it should be less than 1 TeV.\\

If we take
$\Lambda_h$  = 0.35 TeV, the ratio of the mass scales of QHD and QCD would be 1000.
Then we expect new states, e.g. the first excited state of
the weak boson, at about one TeV. At this energy
many narrow resonances should
exist. They would decay into into weak bosons
or into quark or
lepton pairs.
Especially an isospin singlet partner $W^0$ of the Z-boson must exist - it has
the internal structure:
\begin{eqnarray}
W^0 & = & \frac{1}{\sqrt{2}} \left( \overline{\alpha} \alpha +
\overline{\beta} \beta \right) \; .
\end{eqnarray}
The $W^0$-boson would couple to the leptons and quarks like the $Z$-boson.
It must have a mass of at least 0.7 TeV. Otherwise its effects
would have been observed at LEP. It can easily be
produced at the LHC. It would decay into weak bosons,
a Z-boson and a photon or into lepton or quark pairs .\\

If QHD would be identical to QCD, except for the different mass scale,
the $W^0$--boson would have a similar mass as the Z-boson, which is excluded.
The fact that the $W^0$--boson is much heavier than the Z-boson might be related
to a QHD analog of the gluonic anomaly of QCD. The latter implies that the
mass of the $\eta^{\prime}$-meson is much larger than the mass of the neutral pion.
In QHD the isospin singlet axial vector current also has an anomaly, and
this might be the reason why the $W^0$--boson is very heavy.\\

In strong interaction physics there exist excited states of the vector mesons.
Analogously we expect that there exist excited states of the weak bosons.
The excited state of a charged weak boson
can decay into a charged weak boson and a Z-boson or a photon.\\

Below we shall discuss the effects of the QHD interaction for proton-proton
collisions at very high
energy (e.g. for the LHC). The
radius of the proton is of the order of 0.1 fm. At very high
energies the total hadronic cross section is about 60 mb.
The leptons and
quarks have a radius of the order of $10^{-17}$ cm. In the
collisions of two protons at very high energies two quarks can come very
close to each other such that the QHD interaction sets in. The
QHD cross section for quark-quark scattering at very high energies
is expected to be about 60 nb.
In the proton there are three quarks as well as quark-antiquark-pairs and gluons.
The gluons will also
contribute to the QHD cross section. We estimate the QHD cross section in proton-proton collisions
to be of the order of 600 nb.\\

In the collision of two u-quarks a QHD resonance can be produced,
which decays subsequently into weak bosons, lepton pairs or quark pairs. This
resonance can decay into several weak bosons, e.g.
two, three or four Z-bosons, or one Z-boson and a pair of W-bosons.
For example, a neutral spin zero
resonance H with the mass of 1 TeV can decay
into weak bosons:

\begin{eqnarray}
H & \Longrightarrow & Z + W^+ + W^- \; , \nonumber \\
H & \Longrightarrow & Z + Z + Z \; , \nonumber \\
H & \Longrightarrow & Z + Z + W^+ + W^- \; . \nonumber
\end{eqnarray}
\\
Especially interesting are the collisions of quarks and antiquarks.
In these reactions a fermionic haplon can collide with its antiparticle,
or two scalar haplons can collide.
In the collision of two scalar haplons a high mass resonance
can be produced. Likewise
in the collision of
a fermion and an antifermion a resonance can be produced. These resonances
decay into
a lepton or quark pair or into weak bosons. For example, a resonance
with an invariant mass of 1 TeV can in particular decay into three Z-bosons,
into one Z-boson and a pair of W-bosons or
into a pair of muons, electrons, tauons, neutrinos or quarks.
A high mass resonance, decaying into a pair of muons, can easily be detected
with the detectors at the LHC.\\

In a collision of a scalar haplon and a fermionic haplon a very massive fermion
can be produced - an excited lepton or quark, which
should have a mass of the order of 1 TeV. Thus at energies above 1 TeV leptons
become strongly interacting systems. The excited leptons and quarks have
not only spin $1/2$, but also spin $3/2$, etc.
A massive charged lepton resonance would decay into an electron or muon
or tau-lepton or a neutrino, emitting a weak
boson or a photon. For example, an excited muon would decay into a muon, emitting
a Z-boson or a photon. It can also decay into a muon neutrino, emitting a W-boson.\\

We conclude: The weak bosons are interpreted as composite objects,
composed of a pair of basic fermions. The energy scale
of the QHD interaction, responsible for the confinement of the fermions, is estimated to be
just below 1 TeV. Leptons and quarks are composed of a scalar boson and a fermion.
At the energy of 1 TeV a similar complexity arises as in strong interaction
physics at 1 GeV. Many resonances exist above 1 TeV. They should soon
be observed by the detectors at the Large Hadron Collider.

\end{document}